%% file: 5G_Industrial_Network_Architecture.tex
\documentclass[10pt,oneside,english,onecolumn,a4paper]{article}
\usepackage{graphicx}
\usepackage[utf8]{inputenc}
\usepackage{mathptmx}
\usepackage{tabularx}
\usepackage{ragged2e}
\usepackage[singlelinecheck=false]{caption}
\usepackage[T1]{fontenc}
\usepackage[numbers]{natbib}
\usepackage{lscape}
\usepackage{url}
\usepackage{amsmath}
\RequirePackage[blocks]{authblk}
\usepackage{babel}
\usepackage{balance}
\usepackage[nolist]{acronym}
\newcommand\blfootnote[1]{%
  \begingroup
  \renewcommand\thefootnote{}\footnote{#1}%
  \addtocounter{footnote}{-1}%
  \endgroup
}

\makeatletter
\let\ps@plain\ps@empty
\def\@xivpt{14bp}

\def\@sect#1#2#3#4#5#6[#7]#8{%
  \ifnum #2>\c@secnumdepth
    \let\@svsec\@empty
  \else
    \refstepcounter{#1}%
    \protected@edef\@svsec{%
      \ifnum #2<4
        \hb@xt@10mm{\csname the#1\endcsname}\relax
      \else
        \hb@xt@12mm{\csname the#1\endcsname}\relax
      \fi}%
  \fi
  \@tempskipa #5\relax
  \ifdim \@tempskipa>\z@
    \begingroup
      #6{%
        \@hangfrom{\hskip #3\relax\@svsec}%
          \interlinepenalty \@M #8\@@par}%
    \endgroup
    \csname #1mark\endcsname{#7}%
    \addcontentsline{toc}{#1}{%
      \ifnum #2>\c@secnumdepth \else
        \protect\numberline{\csname the#1\endcsname}%
      \fi
      #7}%
  \else
    \def\@svsechd{%
      #6{\hskip #3\relax
      \@svsec #8}%
      \csname #1mark\endcsname{#7}%
      \addcontentsline{toc}{#1}{%
        \ifnum #2>\c@secnumdepth \else
          \protect\numberline{\csname the#1\endcsname}%
        \fi
        #7}}%
  \fi
  \@xsect{#5}}
\renewcommand\LARGE{\@setfontsize\LARGE{16}{20}}
\def\abstract#1{\def\@abstract{#1}}
\def\abstractEn#1{\def\@abstractEn{#1}}
\def\titleEn#1{\def\@titleEn{#1}}
\def\@maketitle{%
  \newpage
  \null
  \let \footnote \thanks
    {\LARGE\bfseries\RaggedRight \@titleEn \par}%
    \vskip 1\baselineskip%
    {\normalsize
      \@author\par}%
    \vskip \baselineskip%
    {\section*{Abstract}
      \@abstractEn}%
  \par
  \vskip 3\baselineskip}

\renewcommand\section{\@startsection {section}{1}{\z@}%
                                   {-3.5ex \@plus -1ex \@minus -.2ex}%
                                   {\baselineskip}%
                                   {\normalfont\Large\bfseries\RaggedRight}}
\renewcommand\subsection{\@startsection{subsection}{2}{\z@}%
                                     {\baselineskip}%
                                     {1ex}%
                                     {\normalfont\large\bfseries\RaggedRight}}
\renewcommand\subsubsection{\@startsection{subsubsection}{3}{\z@}%
                                     {1\baselineskip}%
                                     {3bp}%
                                     {\normalfont\normalsize\bfseries\RaggedRight}}
\renewcommand\paragraph{\@startsection{paragraph}{4}{\z@}%
                                    {1\baselineskip\@plus1ex \@minus.2ex}%
                                    {3bp}%
                                    {\normalfont\normalsize\RaggedRight}}
\renewcommand\subparagraph{\@startsection{subparagraph}{5}{\parindent}%
                                       {3.25ex \@plus1ex \@minus .2ex}%
                                       {-1em}%
                                      {\normalfont\normalsize\bfseries\RaggedRight}}
\parindent\p@
\makeatother
\DeclareCaptionLabelSeparator{enskip}{\enskip}
\titleEn{Towards a Flexible Architecture for Industrial Networking}
\author[1]{Michael Karrenbauer}
\author[1]{Amina Fellan}
\author[1]{Hans D. Schotten}
\author[2]{Henning Buhr}
\author[2]{Savita Seetaraman}
\author[2]{Norbert Niebert}
\author[3]{Stephan Ludwig}
\author[4]{Anne Bernardy}
\author[4]{Vasco Seelmann}
\author[4]{Volker Stich}
\author[5]{Andreas Hoell}
\author[5]{Christian Stimming}
\author[6]{Huanzhuo Wu}
\author[6]{Simon Wunderlich}
\author[6]{Maroua Taghouti}
\author[6]{Frank Fitzek}
\author[7]{Christoph Pallasch}
\author[7]{Nicolai Hoffmann}
\author[7]{Werner Herfs}
\author[8]{Elena Eberhardt}
\author[8]{Thomas Schildknecht}
\affil[1]{Chair for Wireless Communications and Navigation, University of Kaiserslautern, Germany, \protect\\ Email: \{karrenbauer,weinand,schotten\}@eit.uni-kl.de}
\affil[2]{Ericsson GmbH, Herzogenrath, Germany, Email: \{henning.buhr,savita.seetaraman,norbert.niebert\}@ericsson.com}
\affil[3]{Robert Bosch GmbH, Renningen, Germany, Email: \{stephan.ludwig2\}@de.bosch.com} 
\affil[4]{FIR e. V. at RWTH Aachen, Germany, Email: \{anne.bernardy,vasco.seelmann,volker.stich\}@fir.rwth-aachen.de}
\affil[5]{SICK AG, Waldkirch, Germany, Email: \{andreas.hoell,christian.stimming\}@sick.de}
\affil[6]{Deutsche Telekom Chair of Communication Networks, University of Dresden, Germany, \protect\\ Email: \{huanzhuo.wu,simon.wunderlich,maroua.taghouti,frank.fitzek\}@tu-dresden.de}
\affil[7]{Laboratory for Machine Tools and Production Engineering, RWTH Aachen, Germany, \protect\\ Email: \{c.pallasch,n.hoffmann,w.herfs\}@wzl.rwth-aachen.de}
\affil[8]{Schildknecht AG, Murr, Germany, Email: \{elena.eberhardt,thomas.schildknecht\}@schildknecht.ag}

\abstractEn{The digitalization of manufacturing processes is expected to lead to a growing interconnection of production sites, as well as machines, tools and work pieces.
In the course of this development, new use-cases arise which have challenging requirements from a communication technology point of view.
In this paper we propose a communication network architecture for Industry 4.0 applications, which combines new 5G and non-cellular wireless network technologies with existing (wired) fieldbus technologies on the shop floor. 
This architecture includes the possibility to use private and public mobile networks together with local networking technologies to achieve a flexible setup that addresses many different industrial use cases.
It is embedded into the Industrial Internet Reference Architecture and the RAMI4.0 reference architecture.
The paper shows how the advancements introduced around the new 5G mobile technology can fulfill a wide range of industry requirements and thus enable new Industry 4.0 applications.
Since 5G standardization is still ongoing, the proposed architecture is in a first step mainly focusing on new advanced features in the core network, but will be developed further later.}

\begin{document}

\maketitle
\input{./Chapter/Acronyms.tex}
\blfootnote{This is a preprint, the full paper has been accepted by 23th VDE/ITG Conference on Mobile Communication (23. VDE/ITG Fachtagung Mobilkommunikation), Osnabr\"uck, May 2018}
\input{./Chapter/Chapter0-5-Background}

\input{./Chapter/Chapter1-Introduction}
\input{./Chapter/Chapter4-5G}
\input{./Chapter/Chapter5-Architecture}
\input{./Chapter/Chapter5_5-Compatibility}

\input{./Chapter/Chapter6-Conclusion}

\input{./Chapter/Acknowledgement}

\bibliographystyle{unsrtdin}
\bibliography{references/references_v1}

\begin{landscape}
	\begin{figure}
		\centering
			\includegraphics[width=0.9\columnwidth]{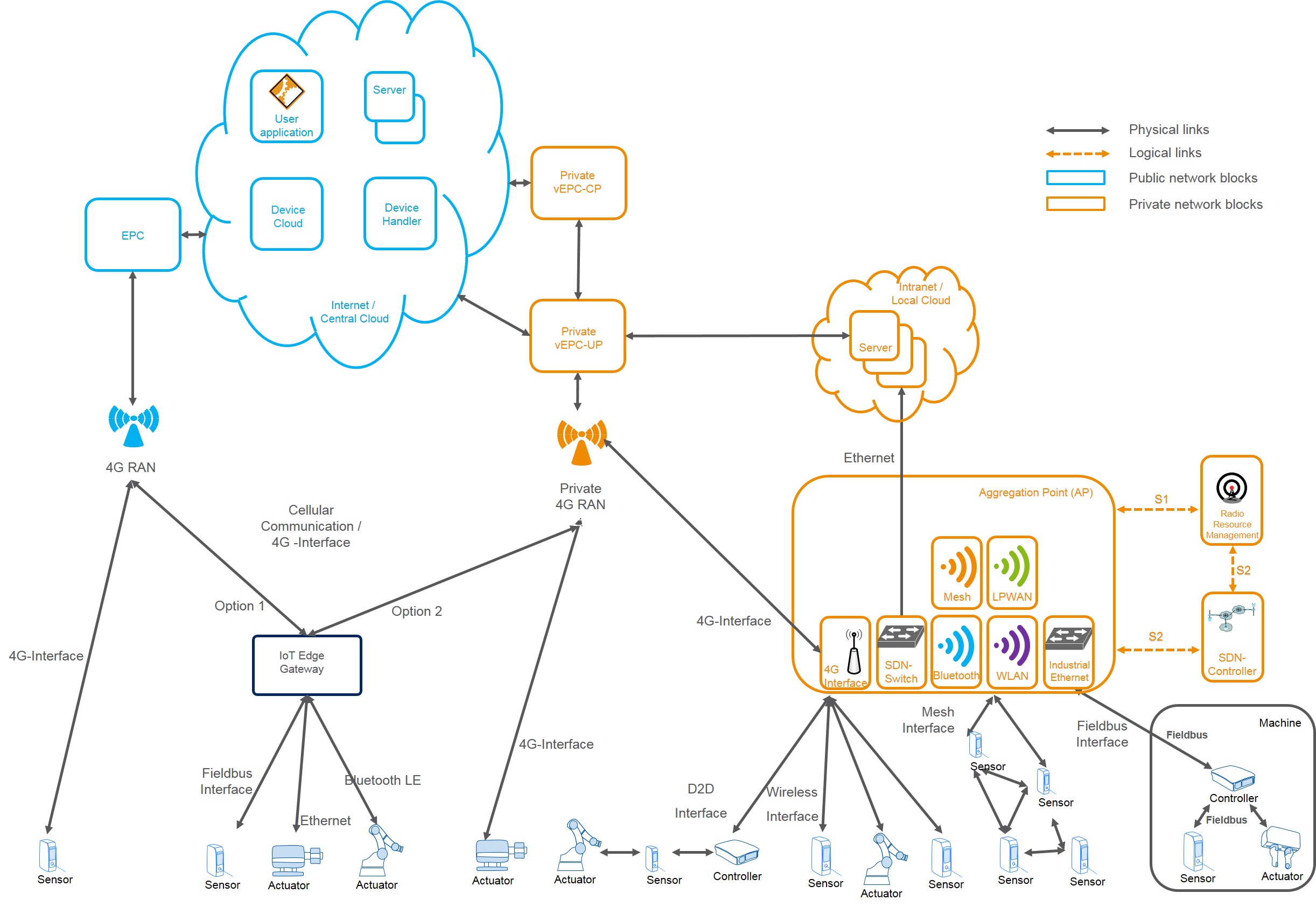}
		\caption{5Gang system architecture}
		\label{fig:architecture}
	\end{figure}
\end{landscape}

\end{document}

%% file: Chapter/Acronyms.tex
\begin{acronym}[MC-SS]
\acro{emb}[EMB]{Enhanced Mobile Broadband}
\acro{iot}[IoT]{Internet of Things}
\acro{itu}[ITU]{International Telecommunication Union}
\acro{iiot}[IIoT]{Industrial Internet of Things}
\acro{mc-ss}[MC-SS]{Multi-Carrier Spread Spectrum}
\acro{mccdma}[MC-CDMA]{Multi-Carrier Code Division Multiple Access}
\acro{mmtc}[MMTC]{Massive machine type communications}
\acro{SDN}[SDN]{Software Defined Network}
\acro{urllc}[URLLC]{Ultra-reliable and low latency communications}
\end{acronym}

%% file: Chapter/Chapter0-5-Background.tex
\section{Background}
The research initiative ``5G: Industrial Internet (5G: II)'' \cite{5GII} of the German Federal Ministry of Education and Research addresses the requirements on a 5G communication network in order to be used with applications of industrial productions. 
The main focus areas are the research on innovative technologies for industrial communication networks, intelligent network management and self-organizing networks. 
As part of this initiative, the 5Gang project \cite{5Gang} considers different use cases of future industrial production principles and their requirements on the communication network.
The project scope does not only cover local production sites but as well the opportunities arising from adding inter-site connections. 
5Gang is a consortium of eight partners from industry and academia who bring in experience covering technical, business and production process aspects.

%% file: Chapter/Chapter1-Introduction.tex
\section{Introduction}
The growing digitalization and interconnection of manufacturing processes is leading to a closer gearing of companies, suppliers and customers.
In the course with this development, advanced communication technologies become necessary.
The intelligent fusion of sensor data enables advanced use cases in industrial environments, like condition monitoring and predictive maintenance, as well as product quality assurance.
In this paper, we describe the way towards a flexible architecture for industrial networking, which is able to support the heterogeneous networking technologies in today's production facilities, while benefiting from the 5G technologies.
The rest of this paper is organized as follows: In Section III, we briefly summarize the industrial communication performance of 5G, which can serve as a basis for our architecture.
In Section IV, the 5Gang architecture and system components are explained in detail.
In Section V, reference architectures known in the literature for Industrial Internet are briefly described and the coverage of our proposed architecture with respect to these reference architectures examined.
A conclusion is drawn in chapter VI.

%% file: Chapter/Chapter4-5G.tex
\section{5G in Production}
The full standardization of 5G is planned to be finished in 2020 but commercial pre-standard systems will be deployed already in 2018.
Requirements might change on its way, but the overall targets are taking shape and can be considered as a base for 5Gang.
Performance is a very important aspect when analyzing the possibilities to use 5G in a production environment.
The United Nations organization ITU-R addresses three main capabilities to define the requirements for 5G systems \cite{ITU}:

\begin{enumerate}
\item Enhanced Mobile BroadBand (eMBB) will provide up to 20 Gbit/s of data towards the end-users (100 Mbit/s per user) and 10 Gbit/s (50 Mbit/s per user) from the end-users towards the network.
The user plane latency shall be below 4 ms.
\item	Ultra-Reliable and Low-Latency Communications (URLLC), sometimes referred to as critical Machine Type Communication (cMTC).
It requires a user plane latency of less than 1 ms.
\item Massive Machine Type Communications (mMTC), will allow to connect up to $1,000,000$ devices per $km^2$ with the given quality of service.
\end{enumerate}
The eMBB and mMTC cases can be covered by today's existing technology, but the URLLC case cannot be reliably supported.
The list below gives some further advantages of 5G networks (not complete):
\begin{itemize}
\item Virtualization of network resources: The same physical network can be used for many use cases.
There is no need to set up and maintain different networks.
Mobile phones of the factory employees can be connected as well.
\item Support for mobility: Moving workpiece carriers can not only be controlled or tracked inside the factory but as well on their way between different production sites.
\item Long battery life: It will be possible to connect battery powered devices in an energy efficient manner allowing operating times of 10 years, in extreme cases even 15 years
\item Privacy: It will be possible to install mobile network equipment in a cost-efficient manner inside the factory.
This will prevent the risk of production parameters or other data leaving the factory site.
\item Security: The usage of SIM cards provides a secure way to manage devices and restrict network access.
\item Economy of scale: The big 5G ecosystem will increase the volume of the radio modems leading to cheaper equipment, which will not be possible if every production solution will lead to extra specialized hardware.
\end{itemize}

Some of the required key technologies are already available today, some other will be developed step by step. They are:
\begin{itemize}
\item Network Slicing: A technique to reserve resources per application and to provide the required quality of service is called Network Slicing \cite{3GPP,3GPP2,Wu2017}.
\item Distributed Cloud (DC): The DC allows to move certain nodes of the mobile network closer to the end-user.
In the case of a production environment this can even mean to deploy not only radio nodes in the factory but core network nodes.
This would not be reasonable with specialized and high-performant mobile nodes but the latest virtualization techniques can bring down the costs.
Instead of moving a complete mobile node, 3GPP has as well standardized the option to move only the user plane closer to the network edge, \cite{3GPP3}.
\end{itemize}

%% file: Chapter/Chapter5-Architecture.tex
\section{5Gang Architecture and System Concept}
\subsection{Components of the 5Gang system architecture}
In this section a detailed overview of the communication architecture components is provided.

\subsubsection{Aggregation Point}
In the core of our proposed communication architecture lies the aggregation point (AP). 
It is primarily responsible for providing connectivity of the networks with large numbers of sensors, controllers, and actuators, all of which could be using different wireless/wired technologies, to the core 5G network. 
The AP is thus to be located at the edge of the network and includes different wireless as well as wired interfaces.
The AP will employ \ac{SDN} solutions to manage the traffic within and between its connected networks.
It comprises of an \ac{SDN} capable switch (e.g., Open vSwitch), to facilitate the traffic management, and an SDN controller (e.g., Ryu, OpenDaylight), to obtain a centralized logical view of the network. 
A radio management system entity is also part of the AP, it manages a large number of connected wireless nodes, while implementing self-organization network techniques to optimize the overall network performance.

\subsubsection{IIoT edge-gateway}
The IIoT edge gateway is designed to be installed on the edge of a network, specifically on or near the data provider and at the edge of the network leading from the machine to the cloud. 
The task of the gateway is to collect, edit and reduce the data available (at the edge), thereby optimizing the transport by the network in terms of speed and costs. 
At the same time, the mobile radio networks and the cloud server are also relieved. It has the following functions:
\begin{itemize}
\item Interface to the process: collection of the operating and diagnostic data of the machine or its sensors. 
For this, numerous inputs are available: The selection ranges from classic 4-20mA/0-10V connection technology, via Bluetooth low energy, Ethernet-based fieldbus (PROFINET, CAN, Modbus, etc.) and future IO-Link up to inputs for Ethernet. 
\item Analysis, pre-processing and caching of the collected data on-site. 
This is ensured by a program that runs in the gateway, which evaluates the data ``intelligently''. 
Additionally, data is compressed and sent to the device cloud in very small---also configurable---data packets. 
\item Interface to the device cloud. 
Processed data is sent via the mobile radio provider with the strongest signal (unsteered roaming).
This data is further directed---possibly via additional networks---and finally via the internet backbone to the specified cloud portal.
\item Low transfer costs. 
The costs depend on the operation mode; the online mode requires permanent connection; the interval mode transmits only regularly at specified times; and the sleep mode in which the device transmits only if needed. 
\item Security: transport, back-end and front-end encryption. 
Transport encryption by server authentication Advanced Encryption Standard (AES) with elliptic-curve Diffie-Hellman (ECDH), device authentication (DAS with ECDH) and hardware authentication (crypto chip); Backend encryption (on the server) through strict isolation of the user databases, access control by means of role-based access control (RBAC) and database encryption using AES; frontend encryption (client) through access control using RBAC and transport layer security (TLS) encryption (https).
\end{itemize}

\subsubsection{Device Cloud}
The device cloud handles tasks such as user, device, and access rights management; sends new settings to the device; or provides users with the ability to further process data using a standardized RESTful API, in ERP systems or data cloud services. 

\subsubsection{Radio Management}
\begin{figure}[!t]
	\centering
		\includegraphics[width=0.7\columnwidth]{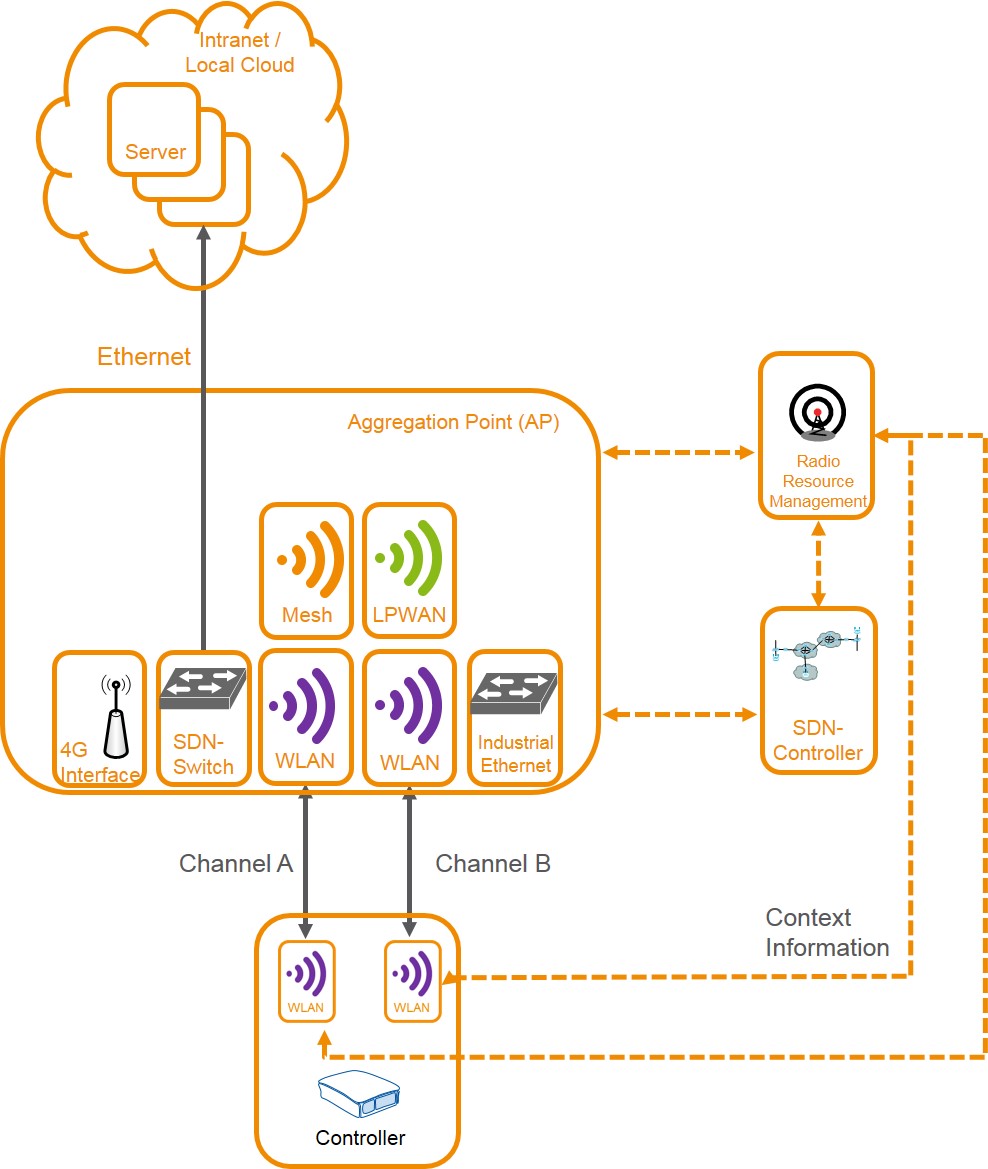}
	\caption{Example for using the ``Radio Management'' Entity}
	\label{fig:radio_management}
\end{figure}

The radio management entity gathers context information about the radio environment and particularly about the link quality of the wireless links associated with the aggregation point. 
This information can be used to apply self-x mechanisms to the overall connectivity solution using the SDN capabilities of the aggregation point. 
One example is shown in Fig.~\ref{fig:radio_management}. 
A device, in this case a ``controller'', is connected to the aggregation point using two independent WiFi adapters. 
In a first step, only one wireless link is established using one WiFi interface which occupies channel A. 
The SDN switch is configured such that traffic going through this interface is routed towards the cloud. 
The second WiFi interface is used for wireless monitoring. 
Both interfaces send context information about the current wireless status to the radio resource management. 
Using this information, if the utilization of a different channel would be beneficial, a second WiFi link is established using channel B.  
The SDN switch is reconfigured, so that the second wireless interface is now routed to the cloud.
After this, the first wireless link can be shut down. 
This process is ``make before break'' and works seamlessly, i.e. there is no packet loss. 
It is also possible to improve this approach with multipath techniques such as Multipath TCP or parallel redundancy protocol (PRP) in a second step.

\subsubsection{SDN Controller}
The SDN controller in an SDN is the centralized management unit of the network, it acts as a strategic controller in the SDN network, manages flow-table on SDN switches, which contains the data activities in the network. 
With the help of an SDN controller, it is possible to configure the network dynamically so that it meets the changing needs related to configuration, security and optimization.
The most popular protocols used by SDN controllers to communicate with SDN switches are OpenFlow and Open vSwitch data-base (OVSDB).
OpenDaylight and Ryu are two typical SDN controllers. 

\subsubsection{SDN Switch}
A SDN switch is a device, which receives, sends and forwards data packets in a network in order to meet specific requirements. 
It follows the rules in a flow-table, which is managed by the SDN controller via SDN protocols, like OpenFlow. 
Dedicated SDN switch hardware is not required, even though many vendors offer specific SDN switches to deliver enhanced SDN performance. 
There are some virtual SDN Switches, e.g. Open vSwitch (OVS), to provide a switching stack for hardware virtualization environments.

\subsubsection{Wireless Sensor Device}
Wireless sensor devices combine a set of sensors, a wireless data transmission technology (connectivity) and a micro controller. 
They are typically powered from an onboard battery in order to be mountable almost everywhere. 
Because of the massive number of devices, e.g. in a production hall, the battery power must last for several years without re-charge. 
Typical sensors are linear acceleration, rotational speed and geomagnetic field orientation, each in 3 axes. 
Typically, these 9 sensor data (9 degrees of freedom, 9DoF) are fused into an high-fidelity orientation value. 
Further sensor data can be light intensity, acoustic (noise)/microphone, humidity, temperature, GPS position, smoke or gas detectors.
Typically, the aggregation point is connected to the Internet by wired or wireless/cellular technologies and connects the sensor to a (cloud) server acting as central instance.
The microcontroller reads the sensors, applies signal processing in order to reduce the data amount and to increase the wireless network efficiency by applying 5G technologies like compressed sensing, distributed source coding or network coding. 
Compressed sensing reduces the sensor reading rate such that a smaller number of samples, albeit still containing all required information, are obtained compared to classic sampling. 
This data is compressed among multiple sensors using distributed source coding. 
Without requiring the exchange of messages between neighboring sensor nodes, the data is encoded on each sensor device independently in such a way that the correlation between different sensors is compressed out of the joint amount of their transmitted data. 
This encoded data is then optionally network encoded over time and then transmitted. 
If the sensor nodes act as a wireless mesh network, neighboring nodes forward the information of other sensors by using network coding. 
After receiving multiple packets of data, either from one or multiple sources, they are network encoded (e.g. by random linear network coding, i.e. linear combinations of packets with randomly chosen coefficients). 
The encoded packets are then broadcasted to other neighboring nodes.

\subsubsection{Anomaly Detection}
An anomaly detection is performed on a central instance, preferably on a (maybe private or edge) cloud server. 
A connectivity middleware, which scales to a massive number of messages and connections, e.g. a MQTT broker and client, is used in order to receive compressed sensor readings from the massive number of sensor devices. 
Before the actual anomaly detection is performed, the compressed signal has to be reconstructed. 
The network coded packet has to be decoded by, in the case of random linear network coding, solving the system of linear equations. 
Then the distributed source decoder has to reconstruct the signal of each source using the encoded data provided by the other sensor devices. 
Finally, the compressed sensed signal has to be converted into a regularly-sampled signal. 
After some pre-processing to the anomaly detection algorithm, this algorithm compares the reconstructed received sensor readings with correct values, which have been learnt in an initial training phase. 
If values deviate too much from the reference, then an alarm is triggered. 
Outlier detection algorithms from machine learning could be used for this purpose.

%% file: Chapter/Chapter5_5-Compatibility.tex
\section{Compatibility with reference architectures}
In the literature, different reference architectures for the Industrial IoT (IIoT) are known.
Two of the most cited architectures are the RAMI 4.0 architecture of the German ``Plattform Industrie 4.0' and the Industrial Internet Reference Architecture (IIRA) of the Industrial Internet Consortium (IIC) which we will check for compatibility with our proposed architecture.

\subsection{RAMI 4.0}
The ``Reference Architectural Model Industrie 4.0'' (RAMI4.0) \cite{Adolphs2015,Hankel2015a} was designed as a future reference model for industrial production and automation to categorize and differentiate different architectural views that are related to each other.
The RAMI4.0 is structured as a three dimensional model comprised of the axes Hierarchy Levels, Layers and Life Cycle Value Stream.
The Hierarchy Levels represent the classical automation pyramid, which structures different layers of responsibility and aggregation from field devices over control hardware to higher-level applications (MES, ERP etc.).
The Hierarchy Levels enhance the classical pyramid by Product and Connected World including the possibility of intelligent (communicating) products as well as interconnecting enterprises or shop floor software technologies to cloud technologies.
The industrial networking architecture presented in this paper covers all hierarchy levels, i.e. its offered functionality addresses every layer in the automation pyramid as well as the connection to superordinated cloud services.
The Layers divide the complete structure into six functional levels.
The Asset layer addresses all physical objects available at production sites including field devices, work pieces, workers etc.
The Integration layer covers all technological methodologies for digitally integrating assets (e.g. attaching QR-Codes on work pieces, using RFID-ID etc.). 
The Communication layer describes the way which communication technology is used for exchanging or accessing data for underlying assets. 
The Information Layer defines how data is represented and transformed into information. 
Furthermore, the Functional Layer describes functionalities for different assets or general system functions. 
Finally, the Business Layer specifies business related information exchange as well as business processes for given production or process segments (e.g. engagement times, down-times, jobs and orders, amount of produced goods etc.). 
The architecture presented in this paper covers the layers starting from ``Asset'' to ``Information''. 
The last dimension of the RAMI4.0 addresses the Life Cycle Value Stream of products. 
It divides the product development and usage process into a type and an instance phase. 
Whereas the type phase refers mainly to product development including documentation, construction plans etc., the instance phase refers to the usage phase of the product, in which data is collected during its operational phase. 
The architecture presented in this paper covers the instance phase of a product lifecycle, which is more challenging from a communication point of view.

\begin{figure}[!t]
	\centering
		\includegraphics[width=0.6\columnwidth]{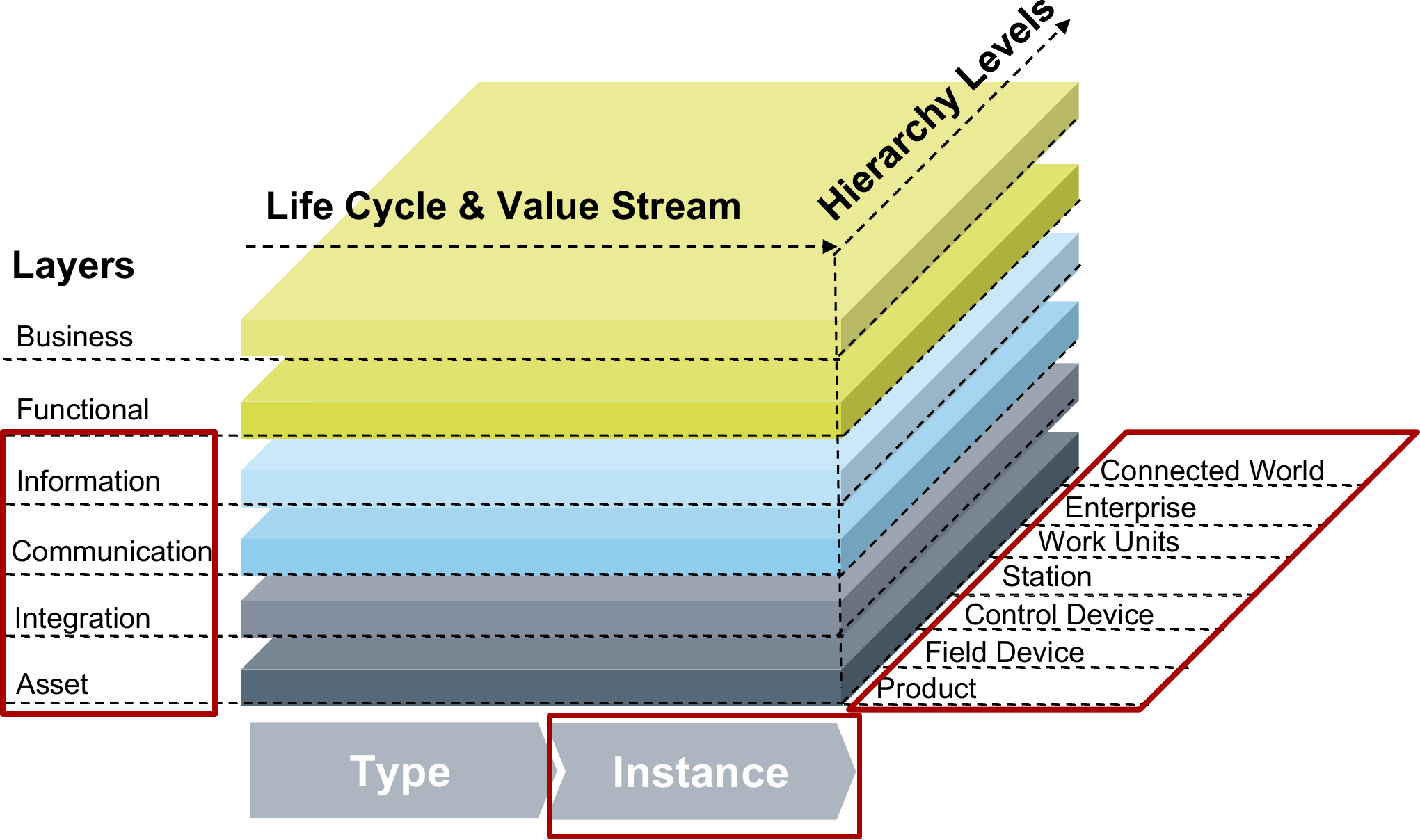}
	\caption{RAMI 4.0 reference architecture}
	\label{fig:rami}
\end{figure}

\subsection{IIRA}
The Industrial Internet Reference Architecture \cite{Lin2015} consists of three dimensions: Functional Domains, System Characteristics and Crosscutting Functions. 
The proposed architecture in this paper covers all functional domains except the ``Business'' domain and all System Characteristics except ``Safety''. 
In terms of crosscutting functions, the proposed architecture is limimted to the functions ``Connectivity'' and ``Distributed Data Management''.

\begin{figure}[!t]
	\centering
		\includegraphics[width=0.5\columnwidth]{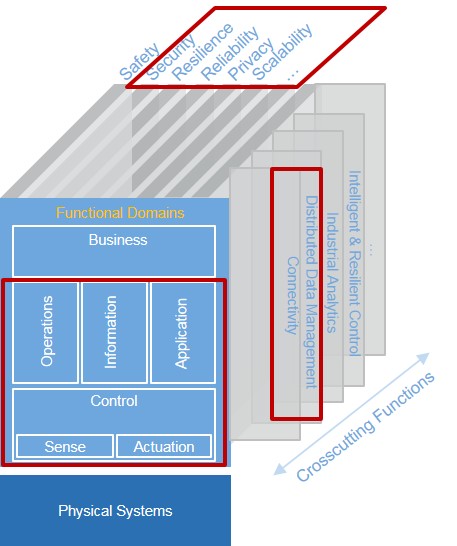}
	\caption{IIRA reference architecture}
	\label{fig:iira}
\end{figure}

%% file: Chapter/Chapter6-Conclusion.tex
\section{Conclusion}
In this paper the authors proposed a communication network architecture for industrial applications combining new 5G and non-cellular network technologies with existing fieldbus technologies on the shop floor, which is able to flexibly address many different industrial use cases. 
Since 5G radio access technologies are still in the standardization process, the proposed architecture focuses on 5G technologies concerning the core network. 
In a second step, the architecture proposed in this paper could be enhanced by 5G radio technologies as soon as they are available.

%% file: Chapter/Acknowledgement.tex
\section*{Acknowledgment}
This work has been supported by the Federal Ministry of Education and Research of the Federal Republic of Germany (Foerderkennzeichen 16KIS0725K, 5Gang). 
The authors alone are responsible for the content of the paper.